\documentstyle[11pt,moriond,epsfig]{article}

\def\Journal#1#2#3#4{{#1} {\bf #2}, #3 (#4)}

\def\PLB{{\em Phys. Lett.}  B}

\newcommand {\ee}        {\rm{e}^+\rm{e}^-}

\newcommand {\pp}        {\rm{p}\bar{\rm{p}}}

\newcommand {\Zzero}     {\rm{Z}}
\newcommand {\MZ}        {m_{\rm{Z}}}
\newcommand {\GZ}        {\Gamma_{\rm{Z}}}
\newcommand {\sigzh}     {\sigma^0_{\rm{h}}}

\newcommand {\Rl}        {R_{\ell}}
\newcommand {\Afb}       {A_{\rm{FB}}}

\newcommand {\Afbzl}     {\Afb^{0,\ell}}
\newcommand {\Afbzb}     {\Afb^{0,\rm{b}}}
\newcommand {\Afbzc}     {\Afb^{0,\rm{c}}}
\newcommand {\Rb}        {R_{\rm{b}}}
\newcommand {\Rc}        {R_{\rm{c}}}

\newcommand {\MW}        {m_{\rm{W}}}

\newcommand {\MT}        {m_{\rm{t}}}
\newcommand {\MH}        {m_{\rm{H}}}

\newcommand {\Ae}        {\cal{A_{\rm{e}}}}

\newcommand {\Atau}      {\cal{A_{\tau}}}
\newcommand {\Al}        {\cal{A_{\ell}}}

\newcommand {\Ab}        {\cal{A_{\rm{b}}}}
\newcommand {\Ac}        {\cal{A_{\rm{c}}}}

\newcommand {\Alr}       {\cal{A_{\rm{LR}}}}
\newcommand {\Alrfb}     {\cal{A_{\rm{FB,LR}}}}

\newcommand {\WpWm}      {\rm{W}^+\rm{W}^-}

\newcommand {\thw}       {\theta_{\rm{W}}}
\newcommand {\alfmz}     {\alpha(\MZ)}
\newcommand {\alfsmz}    {\alpha_s(\MZ)}
\newcommand {\dalfh}     {\Delta\alpha_{\rm{h}}^{(5)}(\MZ)}

\newcommand {\swsqeffl}  {\sin^2\theta_{\rm{eff}}^{\rm{lept}}}
 
\begin{document}
\vspace*{4cm}
\title{ELECTROWEAK DATA AND STANDARD MODEL FIT RESULTS}

\author{ T. KAWAMOTO }

\address{ICEPP, University of Tokyo, \\
7-3-1 Hongo, Bunkyo-ku, Tokyo 113, Japan}

\maketitle\abstracts{
Most recent tests of the Standard Model of the
electroweak interaction are reported using data 
from the four LEP experiments ALEPH, DELPHI, L3 and OPAL, the SLD experiment
at SLC, the Tevatron $\pp$ experiments CDF and D0, and the NuTeV neutrino 
experiment.
Consistency of the Standard Model is studied. 
The value of the Higgs mass is inferred from a global fit.
}

\section{A Global fit and Higgs mass}

\begin{figure}[htb]
\begin{minipage}{0.49\textwidth}
\mbox{\epsfxsize1.0\textwidth\epsffile{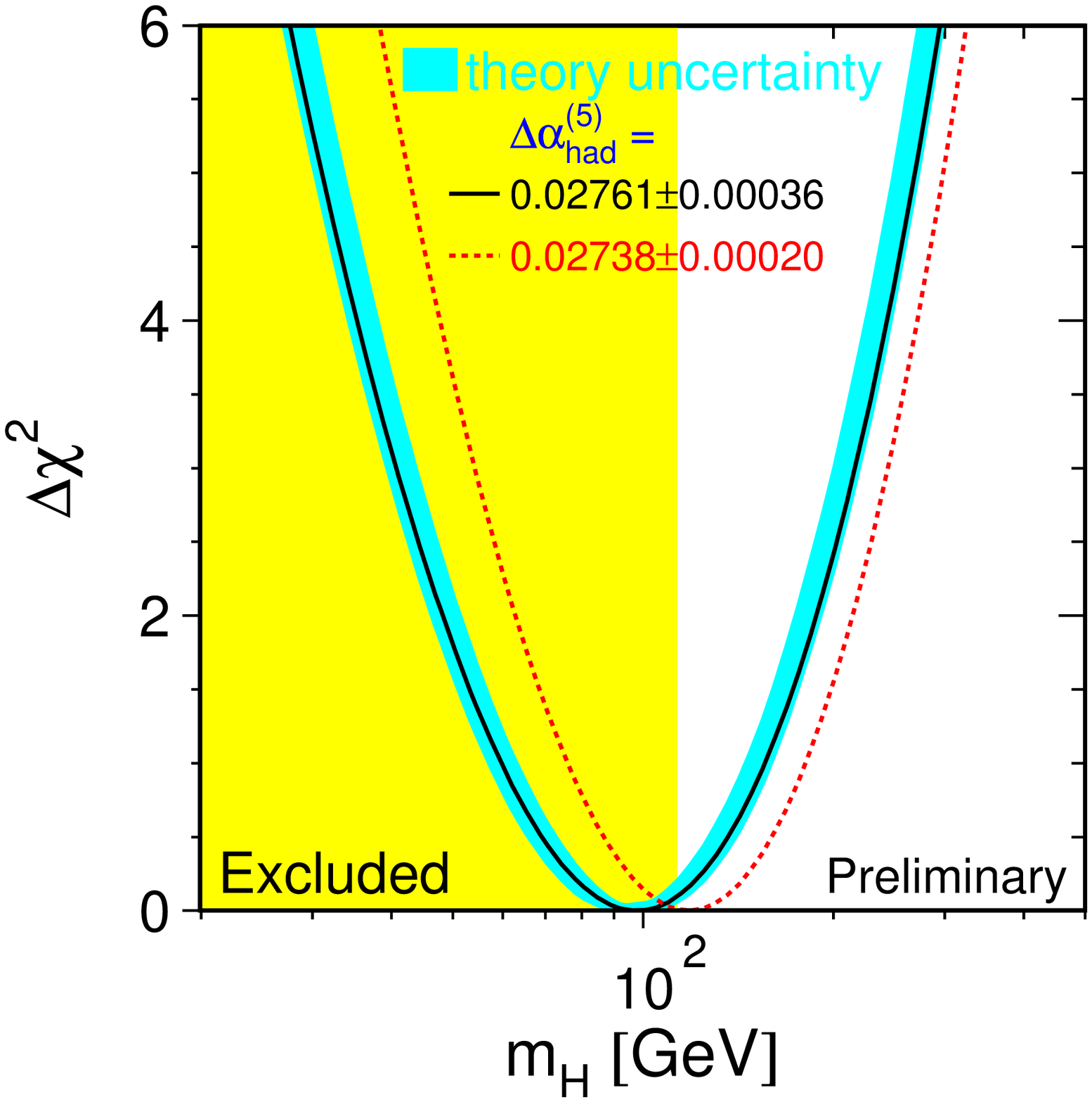}}
\end{minipage}~
\begin{minipage}{0.49\textwidth}
\mbox{\epsfxsize0.8\textwidth\epsffile{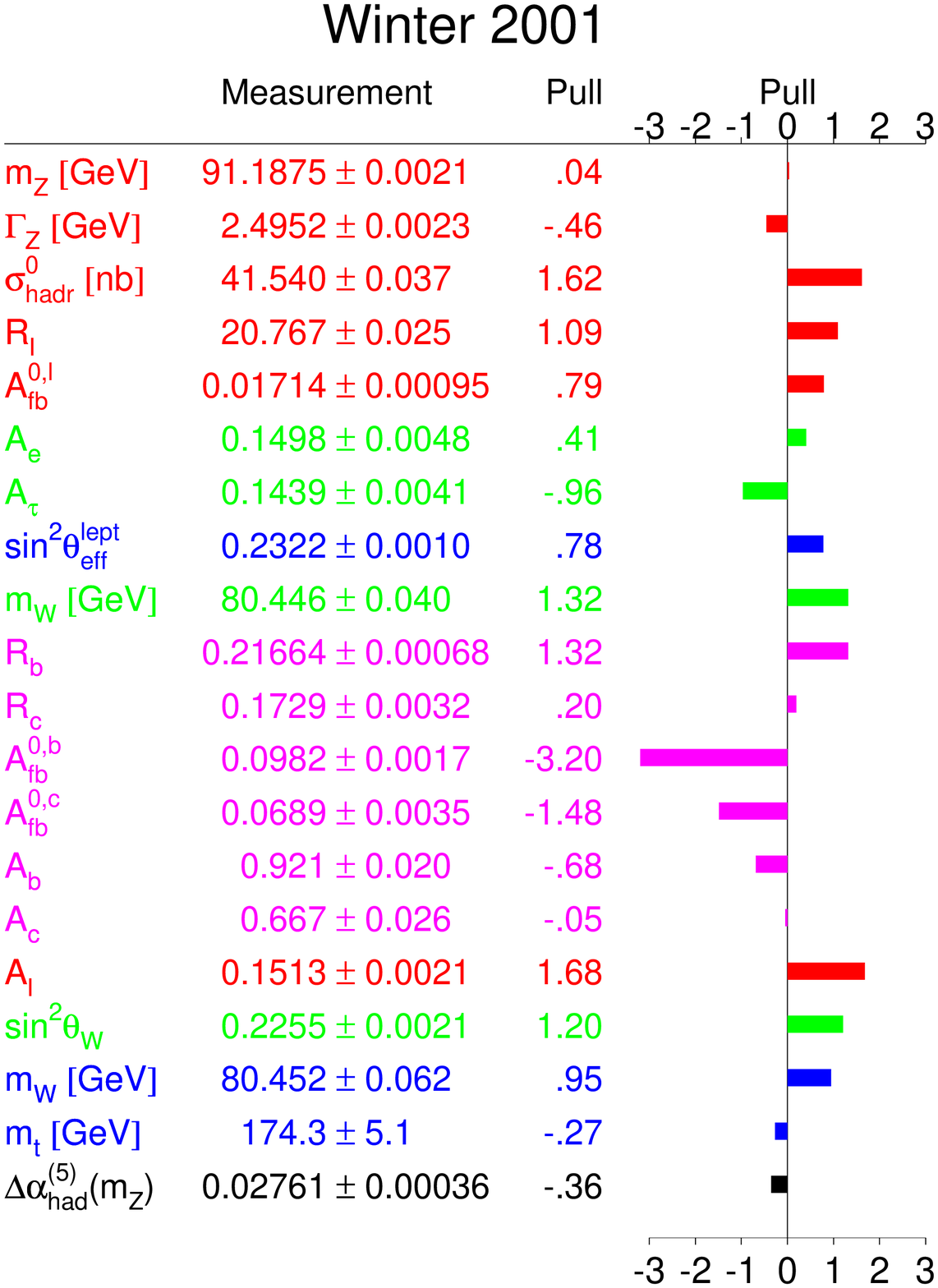}}
\end{minipage}
\caption{Left: $\Delta\chi^2$ of the fit as a function of $\MH$.
The main result is shown by the solid curve.
The associated band represents the estimate of theoretical uncertainty.
The dashed curve is the result using a 
theory-driven determination~\protect\cite{b-martin} of $\dalfh$.
Right: Summary of measurements and the pulls.
\label{f-bb}}
\end{figure}

Latest precision electroweak data (some of them are preliminary) 
are used for a global fit in the framework of the Standard Model (SM).
The data used for the present analysis consist of:
\begin{itemize} 
\item The $\Zzero$ parameters
  \begin{itemize}
  \item lineshape and lepton asymmetry at LEP:
  $\MZ$, $\GZ$, $\sigzh$, $\Rl$ and $\Afbzl$.
  \item $\Ae$ and $\Atau$ from $\tau$ polarisation at LEP. 
  \item $\Al$ from polarised left-right asymmetry by SLD.
  \item Heavy quark (b and c) measurements at LEP and SLD:
  $\Rb$, $\Rc$, $\Afbzb$, $\Afbzc$, $\Ab$, $\Ac$.
  \item $\swsqeffl$ from quark forward-backward asymmetry at LEP.
  \end{itemize}

\item W mass $\MW$ at LEP and $\pp$ colliders.
\item Top mass $\MT$ at $\pp$ collider Tevatron.
\item $\sin^2\thw = 1 - \MW^2/\MZ^2$ from $\nu$N scattering data by the 
      NuTeV experiment.
\end{itemize}
These parameters follow the usual definition~\cite{b-ewwgnote}.
Observables are calculated in the SM with a few input parameters.
$\alfmz$, $\MZ$ and $G_{\mu}$ (the $\mu$ decay constant) are chosen as the 
three basic parameters of the electroweak interaction, and $\alfsmz$ for QCD.
In addition the top quark mass, $\MT$, and Higgs mass, $\MH$ are needed to
calculate higher order corrections.
The latest version of ZFITTER and TOPAZ0 programs are used
\footnote{Recent two loop calculation~\cite{b-deltar} of 
$\Delta r$ is not used here.}.
The parameters in the fits are $\MZ$, $\MT$, $\MH$, $\dalfh$ and $\alfsmz$,
where $\dalfh$ is the light quark contribution to the running of $\alpha$.
Figure~\ref{f-bb} shows the dependence of $\chi^2$ of the fit using 
all data as a function of the Higgs mass $\MH$.
The pulls of data are also shown.
The minimum $\chi^2$ is 25.5/15 d.o.f corresponding to a fit 
probability of 4.3\%.
Data are in general consistent with the SM fit.
An exception is the b forward-backward asymmetry $\Afbzb$ which 
is 3.2$\sigma$ away from the fit.

The result of $\MH$ from the global fit using all data is
\begin{equation}
\MH = 98 ^{+58}_{-38}~~\rm{GeV}~,
\end{equation}
and the upper limit on $\MH$ is 212 GeV at the 95\% CL.
The central value of $\MH$ has increased compared to 
the 2000 summer result~\cite{b-ewwgnote}.
The main reason is due to the update of $\dalfh$ (see below).
When the old value of $\dalfh=0.02804\pm 0.0065$ is used, 
$\MH = 65$ GeV is obtained.

\section{Discussion on the Updates of Electroweak data}

\subsection{Measurements at the $\Zzero$}

All LEP collaborations have finalised the Z lineshape and 
lepton $\Afb$ measurements, and 
the final combination has been made~\cite{b-lscomb}.
Final results of $\Ae$ and $\Atau$ from $\tau$ polarisation measurements
are also available from the LEP collaborations.
A preliminary combination is made.
The SLD collaboration finalised the measurement of $\Al$ using 
polarised left-right cross-section asymmetry, and polarised 
left-right-forward-backward asymmetry $\Alrfb$ for leptonic final states.
Production rate, $\Afb$ and $\Alrfb$ are measured
for b and c quarks from Z decays by the LEP experiments and 
the SLD experiment. 
Combined results of $\Rb^0$, $\Rc^0$, $\Afbzb$, $\Afbzc$, $\Ab$ and $\Ac$ 
are obtained.
New preliminary results of $\Afb^{\rm{b}}$ by ALEPH and DELPHI, and
$\Ab$ and $\Ac$ from $\Alrfb$ and $\Rc$ by SLD are included.

\begin{figure}[htb]
\begin{minipage}{0.49\textwidth}
\mbox{\epsfxsize0.75\textwidth\epsffile{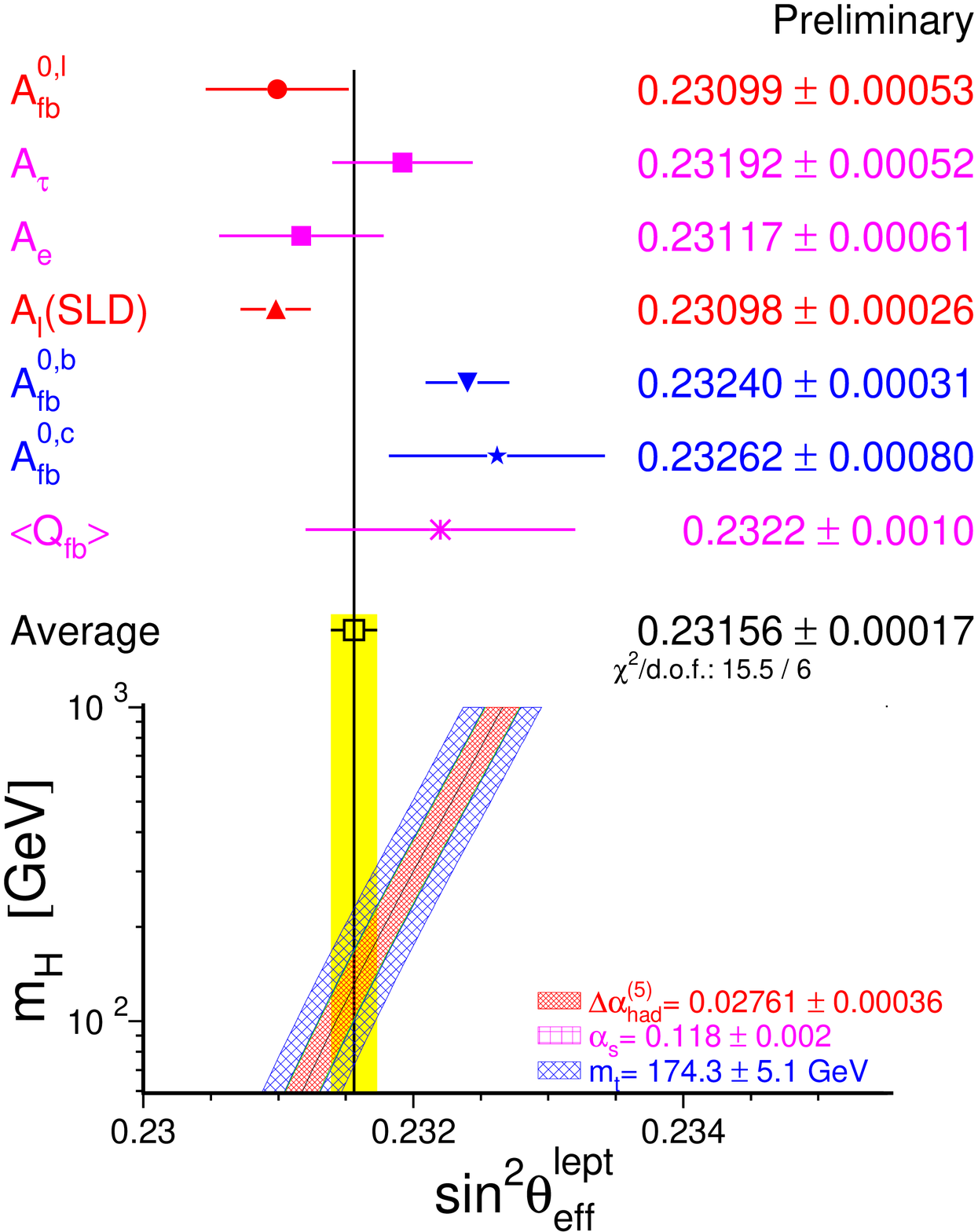}}
\end{minipage}~
\begin{minipage}{0.49\textwidth}
\mbox{\epsfxsize0.95\textwidth\epsffile{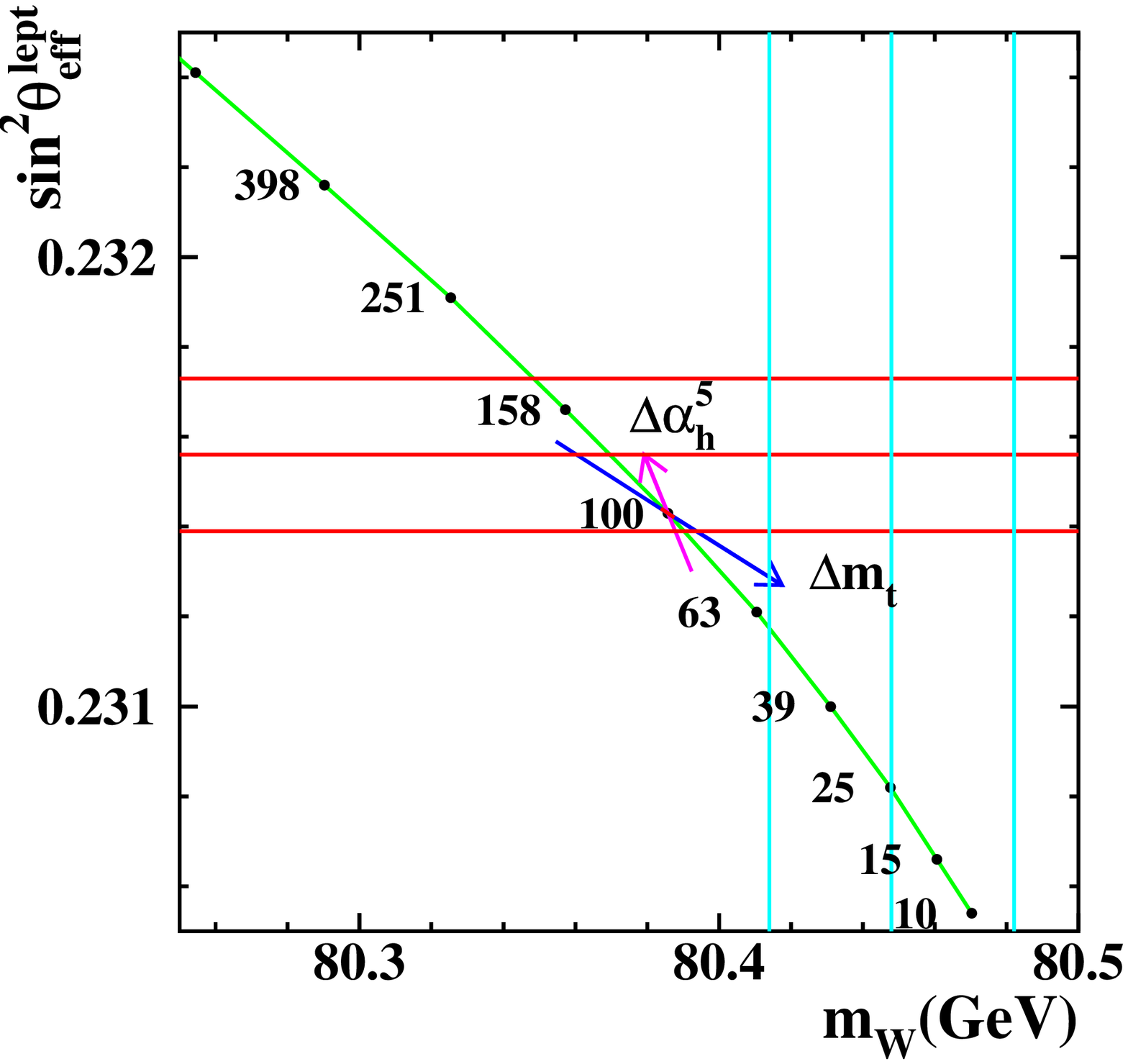}}
\end{minipage}
\caption{Left: Comparison of determinations of $\swsqeffl$ from asymmetry
measurements.  Also shown is the prediction of the SM as a function of $\MH$.
Right: Dependence of $\swsqeffl$ and $\MW$ on $\MH$.  The arrows indicate the
shift of the SM prediction due to uncertainty on $\dalfh$ and $\MT$ for 
$\MH$ around 100 GeV.\label{f-sw2effl}}
\end{figure}

The $\Al$ from the asymmetries can be represented by the
effective electroweak mixing angle $\swsqeffl$.
Figure~\ref{f-sw2effl} shows comparison of results from several asymmetry 
measurements.  
The two most precise ones, from the SLD $\Alr$ and 
the $\Afbzb$ at LEP, show a large difference between them of 3.5 $\sigma$.
This tendency is not new, but due to the reduced
error on $\Afbzb$ the effect becomes sharper.
While the SLD $\Al$ prefers low $\MH$, $\Afbzb$ corresponds to large $\MH$.
%

The parameters $\swsqeffl$ and $\MW$ are sensitive to $\MH$.
Their dependence on $\MH$ is shown in figure~\ref{f-sw2effl}.
The experimental precision of $\swsqeffl$ is similar to the uncertainty
on the SM prediction arising from $\dalfh$.
The uncertainty on $\MT$ corresponds to the precision of both 
$\swsqeffl$ and $\MW$.
Further improvement of $\MT$ and $\dalfh$ would yield significant 
improvements.

\subsection{W mass}

During the period of LEP\,2, the four LEP collaborations have 
collected about 40,000 $\WpWm$ events.
Since the 2000 summer conferences, ALEPH and L3 have updated the 
preliminary $\MW$ results including the data from year 2000.
The combined preliminary $\MW$ result from LEP is:
\begin{equation}\label{e-mw1}
\MW = 80.446 \pm 0.026 (\rm{stat}) \pm 0.030 (\rm{sys})~~(\rm{LEP ~prelim.}).
\end{equation}
This and the result from $\pp$ colliders, $\MW = 80.452 \pm 0.062~~(\pp)$,
are in good agreement.
A current average is :
$\MW = 80.448 \pm 0.034~~(\pp + \rm{LEP})$.

\begin{figure}[htb]
\begin{minipage}{0.49\textwidth}
\mbox{\epsfxsize0.8\textwidth\epsffile{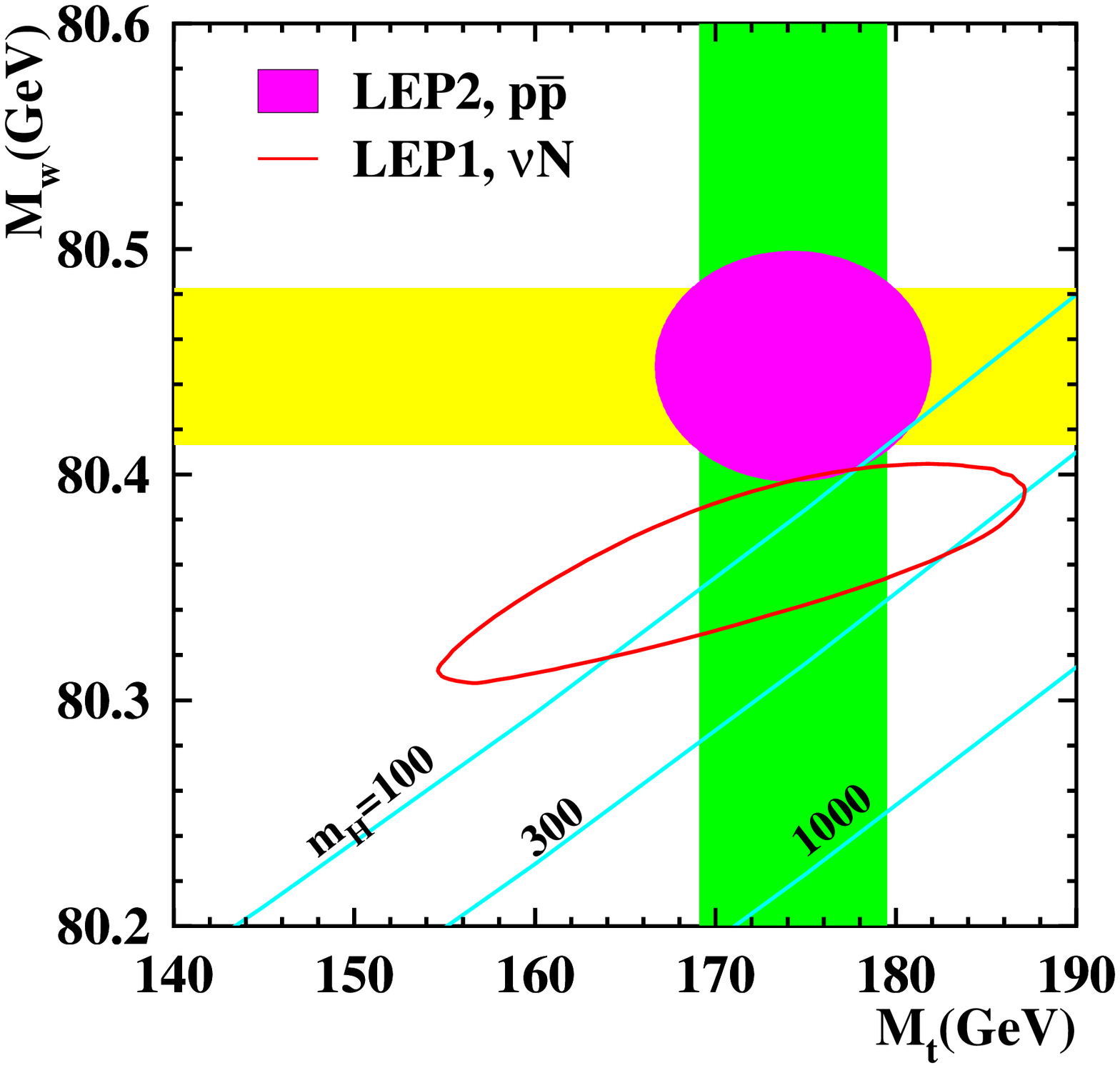}}
\end{minipage}~
\begin{minipage}{0.49\textwidth}
\mbox{\epsfxsize0.8\textwidth\epsffile{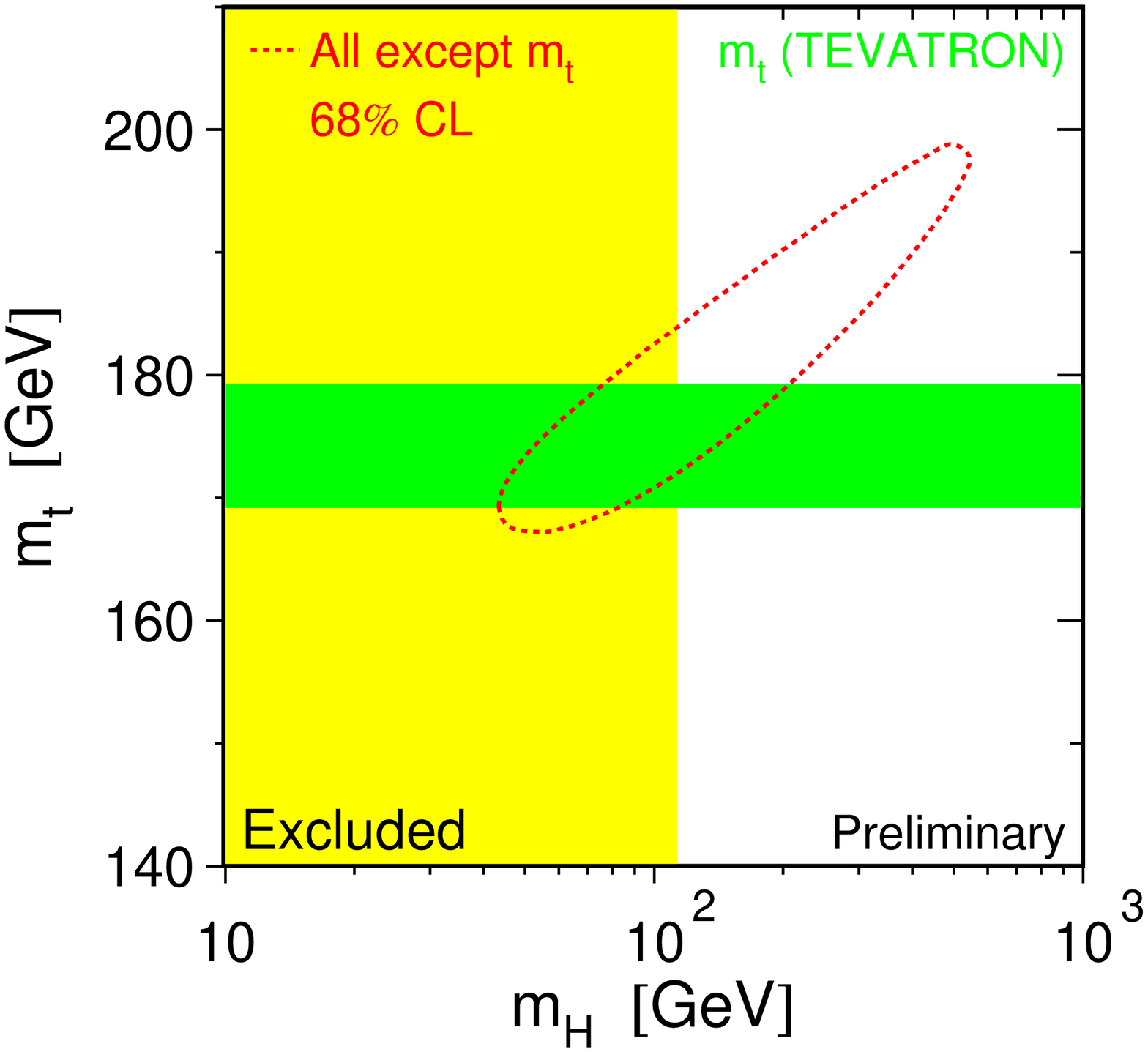}}
\end{minipage}
\caption{Left: The 68\% CL contours in the $\MW$-$\MT$ plane for
the direct and indirect determinations.  
Right: The 68\% CL contour in the $\MT$-$\MH$ plane and comparison with 
the direct $\MT$ determination.
\label{f-mwmt}}
\end{figure}

The direct measurement of $\MW$ can be compared to indirect determination
from the SM fit using the Z and $\nu$N data.
Figure~\ref{f-mwmt} makes this comparison in the $\MW$ - $\MT$ plane.
Also shown in figure~\ref{f-mwmt} is a comparison of direct $\MT$ 
measurement and the result of the SM fit without using $\MT$.
These are in agreement each other.

\subsection{$\alfmz$}

Though the QED coupling constant $\alpha(0)$ is precisely known, 
evaluation of $\alfmz$ at the energy scale $\MZ$ requires elaboration.
While contributions from leptonic loops have been accurately calculated,
the largest uncertainty is on the contribution from light quarks.
The hadronic contribution $\dalfh$ is calculated by a dispersion integral
of the total $\ee$ hadronic cross-section, 
$R=\sigma_{\rm{had}}/\sigma_{\mu\mu}$. 
The precision is limited by the error on the $R$ data, in particular at 
low energies. 
Attempts have been made to improve the accuracy by 
adopting QCD at low energy~\cite{b-martin}.
The BES experiment recently reported a new measurement
of $R$ in the energy range 2 - 5 GeV \cite{b-bes}.
This allows improved determination of $\dalfh$ not relying on QCD.
A new evaluation~\cite{b-handb} using the BES result
is used in the analyses presented here.
\begin{equation}\label{e-dalf}
\dalfh = 0.02761 \pm 0.0036.
\end{equation}

\section{Summary}

The combined results of many precision electroweak measurements allows
tests of the Standard Model. 
The results are sensitive to $\MH$ and an estimate 
is obtained from a global fit.
Most of the data are consistent with each other and 
agree with the SM predictions.
A large difference in the results interpreted in terms of $\swsqeffl$ 
is observed.
In the global SM fit, this manifests itself as a large deviation of $\Afbzb$.
The indirect $\MW$ from the SM fit also shows a slight 
deviation from the direct measurement.
There are many electroweak data still to be finalised.
Thanks are due to the LEP electroweak working group and 
all those who helped me.

\end{document}